\documentclass[pra,10pt,twocolumn,superscriptaddress,showpacs]{revtex4-2}
\usepackage[pdftex, pdftitle={Article}, pdfauthor={Author}]{hyperref}
\usepackage{epsfig}
\usepackage{amsmath}
\usepackage{graphicx}
\usepackage{graphics}
\usepackage{tikz}
\usepackage{float}
\usepackage{amssymb}
\usepackage{tikz}
\usepackage{natbib}
\usepackage{epstopdf}
\usepackage{wrapfig}
\usepackage{mathtools}
\usepackage{xcolor}
\RequirePackage{tcolorbox}
\RequirePackage[outline]{contour}
\usepackage{amsfonts}
\usepackage{dsfont}
\usepackage{bm}
\usepackage{physics}
\usepackage{verbatim}

\renewcommand{\ket}[1]{\left| #1\right\rangle}

\usepackage[english]{babel}

\usepackage{hyperref}
 \hypersetup{
     colorlinks=true,
     linkcolor=blue,
     filecolor=magenta,      
     urlcolor=blue,
     citecolor=blue}
\usepackage{empheq}

\definecolor{fore}{RGB}{249,242,215}
\definecolor{myblue}{rgb}{.8, .8, 1}
\definecolor{forshading}{RGB}{185,217,255}
\newcommand*{\boxedcolor}{blue}
\renewcommand{\boxed}[1]{\textcolor{\boxedcolor}{\fbox{\normalcolor\m@th$\displaystyle#1$}}}
\makeatother

\begin{document}

\title{AC nuclear Stark effect in H-atom via super-intense laser-atom interaction}
\author{Ali Raza Mirza}
\email{a.r.mirza@surrey.ac.uk}
\affiliation{Department of Mathematics and Physics, University of Surrey, GU2 7XH, Guildford, United Kingdom}
\affiliation{Department of Physics, Government College University Lahore, Katchery Road, Lahore 54000, Pakistan}

\author{Rizwan Abbas}
\affiliation{Department of Physics, Lahore University of Management Sciences, Lahore 54792, Pakistan}

\author{Atif Shahbaz}
\affiliation{Department of Physics, Government College University Lahore, Katchery Road, Lahore 54000, Pakistan}

\begin{abstract}
    We investigate the nuclear Stark effect induced in hydrogen-like atomic nuclei under super-intense laser fields. Since laser wavelengths are generally larger than nuclear dimensions, direct laser-nucleus interaction is unfeasible. Instead, this effect is induced indirectly through electron oscillations in the laser field, which produce a periodic electric field that shifts the nuclear energy levels. Using perturbation theory, we derive an expression for the energy shift and dynamic polarizability of the nucleus as a function of laser parameters. Our findings reveal that the Nuclear Stark effect can be controlled by adjusting the laser frequency and intensity, potentially enabling applications in nuclear and quantum optical systems.
\end{abstract}

\maketitle
\section{Introduction} \label{sec1}
An atomic ionization is one of the most fundamental quantum mechanical phenomena and its understanding has implication for a variety of physics topics. In the last three decades, electric field ionization of atoms has attracted a great deal of experimental and theoretical interest \cite{a1,a2,a3,a4,a5}. The interaction between laser light and atoms is a fundamental area of study in atomic physics, marked by high intensity, coherence, monochromaticity, and directionality, enabling unique nonlinear atomic and molecular processes which are unattainable with ordinary light. Perturbation theory has been extensively used for such theoretical investigations \cite{a6,a7,a8,a9,a10,a11,a12}. When a hydrogen-like atom is subjected to an external uniform electric field (constant field), its spectral lines split, a phenomenon known as the Stark effect\cite{a6}. However, the phenomenon of AC shifting of atomic energy levels under a laser field instead of a static electric field, known as the AC Stark shift\cite{a13}. Multi-photon ionization spectroscopy is one of the primary methods used to observe and quantify these AC Stark shifts, as it enables the detection of ionization rates influenced by laser-induced energy level shifts\cite{a15}. The study of energy level shifts under external fields has revealed critical insights into both atomic and nuclear systems. While the Stark effect has been extensively studied in atomic and molecular systems, recent interest has extended this phenomenon to the nuclear domain. Likewise, the nucleus being a bound state has quantized energy levels \cite{heyde1994nuclear}. Shift in these levels under the influence of an external electric field, resulting in what is known as the nuclear Stark shift\cite{a16,a17}. However, when the electric field is substituted with a laser field, a new, time-dependent effect known as the AC nuclear Stark shift. This effect emerges via indirect laser-nuclear interaction. For this purpose, we employ a super-intense laser of which we study the splitting in the energy spectrum of the hydrogen atom.

The challenge over here is that the direct probing of nuclear structures using laser fields is limited by the wavelength of currently available laser technologies. This inherent wavelength limitation prevents direct investigation of nuclear structures. To address this, we propose an alternative technique that allows for the indirect probing of the nucleus. This approach, enabling detailed studies of nuclear properties beyond the constraints imposed by current laser wavelength limitations. To this end, increasing the intensity of the laser field for interaction leads to additional phenomena, such as pair production, vacuum polarization, and the Schwinger limit\cite{a18,a19,a20,a21}, which do not apply direct interaction. In our scheme, when a hydrogen-atom is subjected to a polarized laser field, the electron starts oscillating along the direction of the polarized field. This oscillation effectively influences the nuclear energy levels, which are responsible for the AC nuclear Stark shift. We use perturbation theory to calculate AC nuclear Stark shift analytically. 

Our first goal here is to calculate analytically the Stark shift in the energy levels of the Hydrogen atom nucleus under the action of super-intense laser. This paper is organized as follows. In section ~\ref{sec2}, we introduce our intuitions and present detailed analytical derivation of nuclear Stark shift due to time varying electric field of laser needed for further investigations. In the next sec. ~\ref{sec3}, we show graphical results with brief explanation. We investigate the response of the Stark shift by varying laser parameters such as frequency, electric field strength and time. Lastly, we summarize our findings in section \ref{sec4}.

\section{AC nuclear Stark shift} \label{sec2}

We are concerned with a one-electron atom (or Hydrogen-like ion) with a point-like, spinless and motionless nucleus is placed in uniform electric field of $10^6 V/cm$, the shift in energy levels due to the external electric field is large compared to fine structure, in the strong field limit, the Stark effect is independent of electron spin. A uniform electric field shifts the atomic energy levels, and the law of conservation of energy is valid only, but this is not true in a time-dependent field. Now the profile of applied external field along the z-axis is $\mathcal{\vec{E}}(t) = \mathcal{E}_0 \cos(\omega t)\hat{z}$, where  $\mathcal{E}_0$ is the amplitude of the applied electric field. The interacting Hamiltonian is $H'(t) = -\mathcal{E}_0 D_{z} \cos(\omega t)$. In the presence of a time-varying electric field, $\psi(x, t)$ satisfies the time-dependent Schrödinger Equation:

\begin{align}
    (H_0 + H') \ket{\psi(x, t)}
    =i\hbar \partial_t |\psi(x,t)\rangle.
\end{align}
The wave function $|\psi(x,t)\rangle$ can be written in a superposition of a number of stationary states $|\psi(x)\rangle$\cite{a23}.
\begin{align*}
    |\psi(x,t)\rangle 
    = \sum_k C_k |\psi(x)\rangle e^{-iE_k t / \hbar}.
\end{align*}
Suppose we switch on the electric field at $t = 0$. Moreover, let the atom be in the state $a = \gamma', M_j$. The initial condition is $C_k(t) = \delta_{ka} \quad \text{for} \quad t \leq 0$. Now the amplitude is define as   $C_a(t)=|C_a(t)| e^{-i\eta(t)}$, where $\eta$ is a real phase and $\eta(0) = 0$. The probability of finding the atom in the $k$-state at any time “$t$” is $|C_a(t)|^2$, which is very small for $k \ne a$ unless the resonance occurs at $\omega = |\omega_{ka}|$, but here we only study the case for which $\omega \ne |\omega_{ka}|$, so that $|C_k(t)| \ll 1$ for $k \ne a$ and $C_a(t) \simeq 1$. To understand the significance of the phase $\eta(t)$, we note that the term for which $k = a$ in the expansion
\begin{align*}
    C_a(t)|\psi_a(x)\rangle e^{-i E_a t / \hbar} 
    &= |C_a(t)||\psi_a(x)\rangle
\\
    &\times \text{Exp}\left[-i/\hbar \int_0^t \{E_a + \Delta E_a(t')\} dt'\right],
\end{align*}
with $\Delta E_a(t') = \hbar \dot{\eta}(t')$ and $\dot{C}_a(t) = \left(|\dot{C}_a(t)|-i C_a(t)\dot{\eta}(t)\right )e^{-i\eta(t)}$. Plugging all these values into the  Schrödinger equation, we get,
\begin{align*}
    \dot{C}_b(t) 
    = -\frac{i}{\hbar} \sum_k H'_{bk}(t) C_k(t) e^{i\omega_{bk}t'}.
\end{align*}
These sets of coupled equations can be solved to second-order perturbation $H'(t)$ to get
\begin{align*}
    C_a(t) 
    \simeq 1 + (i\hbar)^{-2} \sum_{k \ne a} \int_0^t H'_{ak}(t') e^{i\omega_{ak} t'} dt' 
\\
    \times \int_0^{t'} H'_{ka}(t'') e^{i\omega_{ak} t''} dt''.
\end{align*}
Now
\begin{align*}
    \dot{C}_a(t) 
    = (i\hbar)^{-2} \sum_{k \ne a} e^{i\omega_{ak} t} \int_0^t H'_{ak}(t') e^{i\omega_{ka} t'} dt',
\end{align*}
where $H'_{aa}(t) = 0$. We note that the unperturbed amplitude $C_a^{(0)} = 1$ and   $\dot{C}_a(t) \approx i\eta(t)$,so the energy become as 
\begin{align*}
    \Delta E_a 
    = (i\hbar)^{-1} \sum_{k \ne a} H'_{ak}(t) e^{-i\omega_{ak} t} \int_0^t e^{-i\omega_{ak} t'} H'_{ka}(t') dt'.
\end{align*}
We are interested in the quantity $\Delta\overline{ E}_a$, which is the mean value of $\Delta E_a(t)$, i.e.
\begin{align}
    \Delta\overline{ E}_a 
    = -\frac{1}{2} \mathbb{E}_{\text{av}}^2(t) \mu(\gamma', M_j, \omega), \label{eq1}
\end{align}
where $\mu(\gamma', M_j, \omega)$ is the dynamic polarizability, define as 
\begin{align*}
    \mu (\gamma', M_j, \omega) 
    = 2\sum_{\gamma'J'} \frac{(E_{\gamma' J'} - E_{\gamma J}) |\langle\gamma'J'M_j|D_z|\gamma J M_j \rangle |^2}{(E_{\gamma' J'} - E_{\gamma J})^2 - \hbar^2 \omega^2}.
\end{align*}
We want to study the changes in these energy levels as it is subjected to a laser. First we need to study the motion of an electron under the action of a laser of which the profile of electric field is $\mathcal{E}(t)=\mathcal{E}_0e^{-i\omega t}$. Thus it is straightforward to calculate the oscillating amplitude of an electron in the presence of such field \cite{a24}.
\begin{align*}
    \alpha(\omega, t) 
    = \frac{e \mathcal{E}_0 \sin \omega t}{m (\omega_0^2 - \omega^2)}.     
\end{align*}
Here $\omega$ is the frequency of the laser and $\omega_0$ is the frequency of oscillating electron. Due to this electron's oscillation, the nucleus experiences a periodic electric field. We are interested in the average field $\mathbb{E}_{\text{av}}$ experienced by the nucleus. When the electron is at maximum distance from the nucleus
\begin{align*}
    \mathbb{E}_1(\omega, t)
    = \frac{-ke}{\{r_0 + \alpha(\omega, t)\}^2},
\end{align*}
on the hand, when an electron is close to the nucleus, we have
\begin{align*}
    \mathbb{E}_2(\omega, t)
    = \frac{-ke}{\{r_0 - \alpha(\omega, t)\}^2}.
\end{align*}
Thus, the average electric field calculated as
\begin{align*}
    \mathbb{E}_{\text{av}}(\omega, t) 
    = -\frac{ke}{2} \left[ \frac{1}{\{r_0 - \alpha(\omega, t)\}^2} + \frac{1}{\{r_0 + \alpha(\omega, t)\}^2} \right],
\end{align*}
where $r_0$ is the Bohr radius, the expectation value $\langle\gamma'J'M_j|D_z|\gamma J M_j \rangle=er_0$ and within atomic units, we set $k=m=\hbar=e=r_0=1$ for convenience. Although we have everything that we need, nevertheless, it is important to look into the restrictions on the choice of laser parameters. First, the necessary condition to avoid electron-nucleus collision, amplitude of oscillation $\alpha(\omega, t)$ of the electron less than the Bohr radius $r_0$ that is  $\frac{e \mathcal{E}_0 \sin \omega t}{m (\omega_0^2 - \omega^2)} < 1$ which means to prevent system to be ionized the $\omega < 0.499 a.u$ and another important factor we must take care of non-relativistic effect, $q = \frac{\mathcal{E}_0}{emc} < 1$, We may ignore all relativistic effects in the motion of electron. The Stark splitting for two nuclear states of H-atom is $E_{\gamma' J'} - E_{\gamma J} = 2.94 \times 10^4 \text{ a.u.}$\cite{a25}. Now we are equipped to use Eq. \eqref{eq1} and investigate the Stark splitting.

\section{Results} \label{sec3}
\begin{figure}[h]
    \centering
    \includegraphics[width=0.45\textwidth]{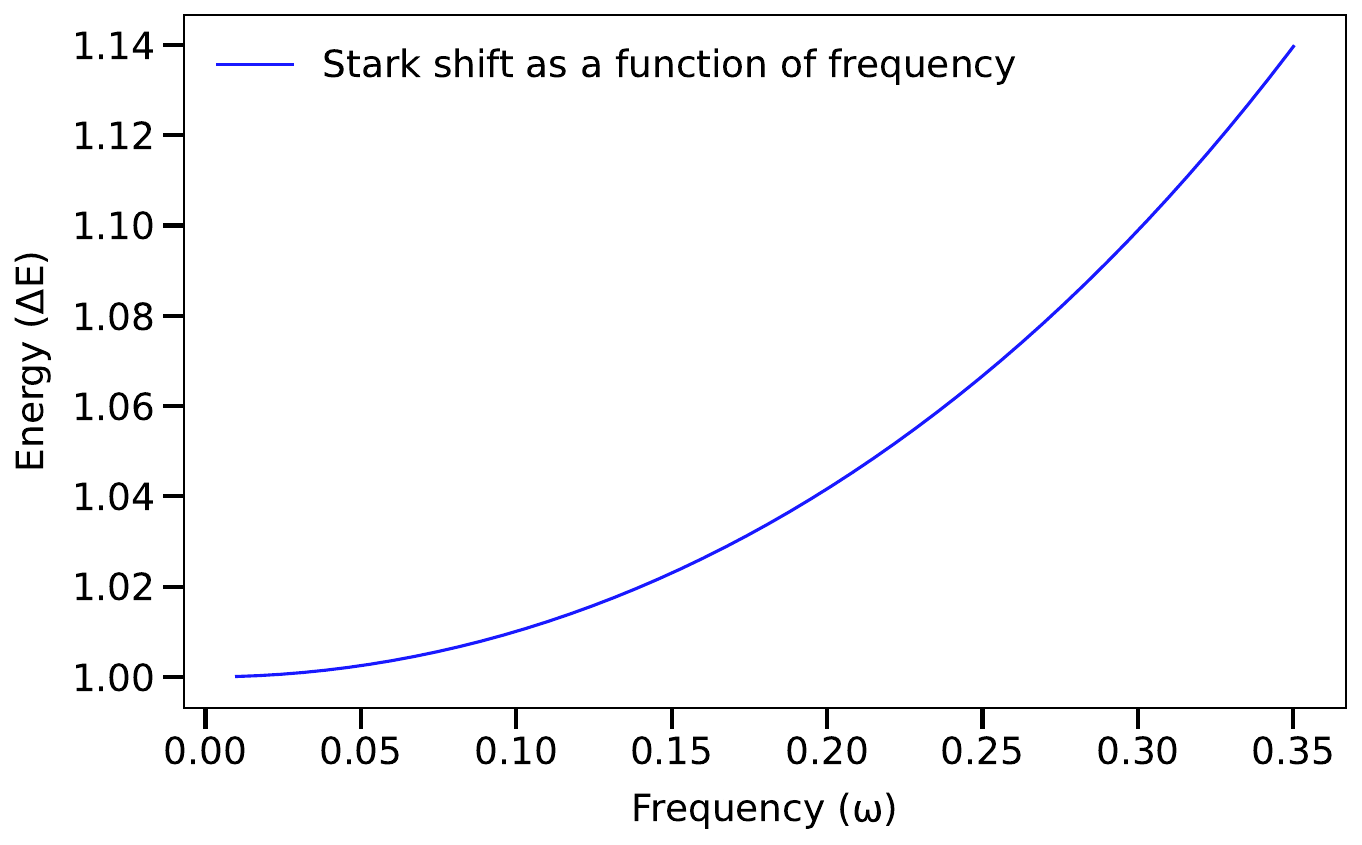}
  \caption{Behavior of Stark effect as a function of time. We choose laser parameters $\mathcal{E}=0.1  \text{a.u}$ and $t=1 \text{a.u}$.}
\label{fig1}
\end{figure}

We carefully choose the parameters of the external laser to avoid ionization or electron-nucleus collision, etc. We have already discussed these restrictions in detail in the previous section. We now use Eq. \eqref{eq1} to investigate the behavior of Stark splitting as a function of laser frequency $\omega$, intensity $E$ and time. As the frequency of the external laser is raised, the oscillating frequency of the electron of H-like atom also increases. As a result, the nucleus experiences the electric field for a longer time, thus we can expect continuous increase in energy splitting. This is precisely what is illustrated in Fig. \ref{fig1} where Stark shift increases with laser frequency. As alluded earlier, when we increase frequency, the number of interactions per second of an electron with the nucleus increases, hence the electric field per second experienced by the nucleus increases, due to which shift in the energy levels in the H-atom nucleus also increases. In other words, the electron remains in contact with the nucleus for a greater time, making this a continuous rise in the Stark shift.

\begin{figure}[H]
  \centering
  \includegraphics[width=0.45\textwidth]{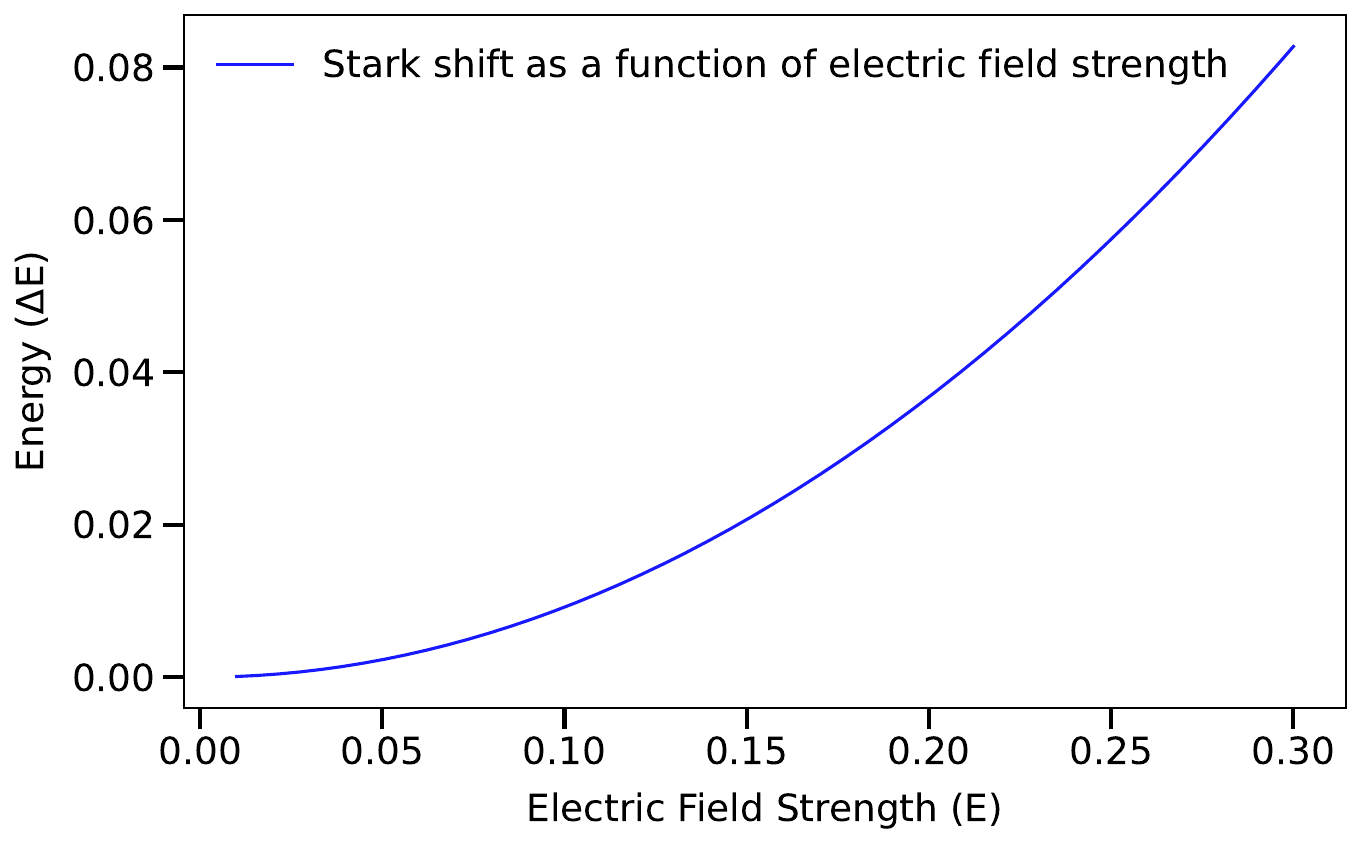}
  \caption{Behavior of Stark shift as a function of intensity of external laser. We choose laser parameters $\omega=0.1  \text{a.u}$ and $t=10^2 \text{a.u}$.}
\label{fig2}
\end{figure}

\begin{figure}[h]
  \centering
  \includegraphics[width=0.45\textwidth]{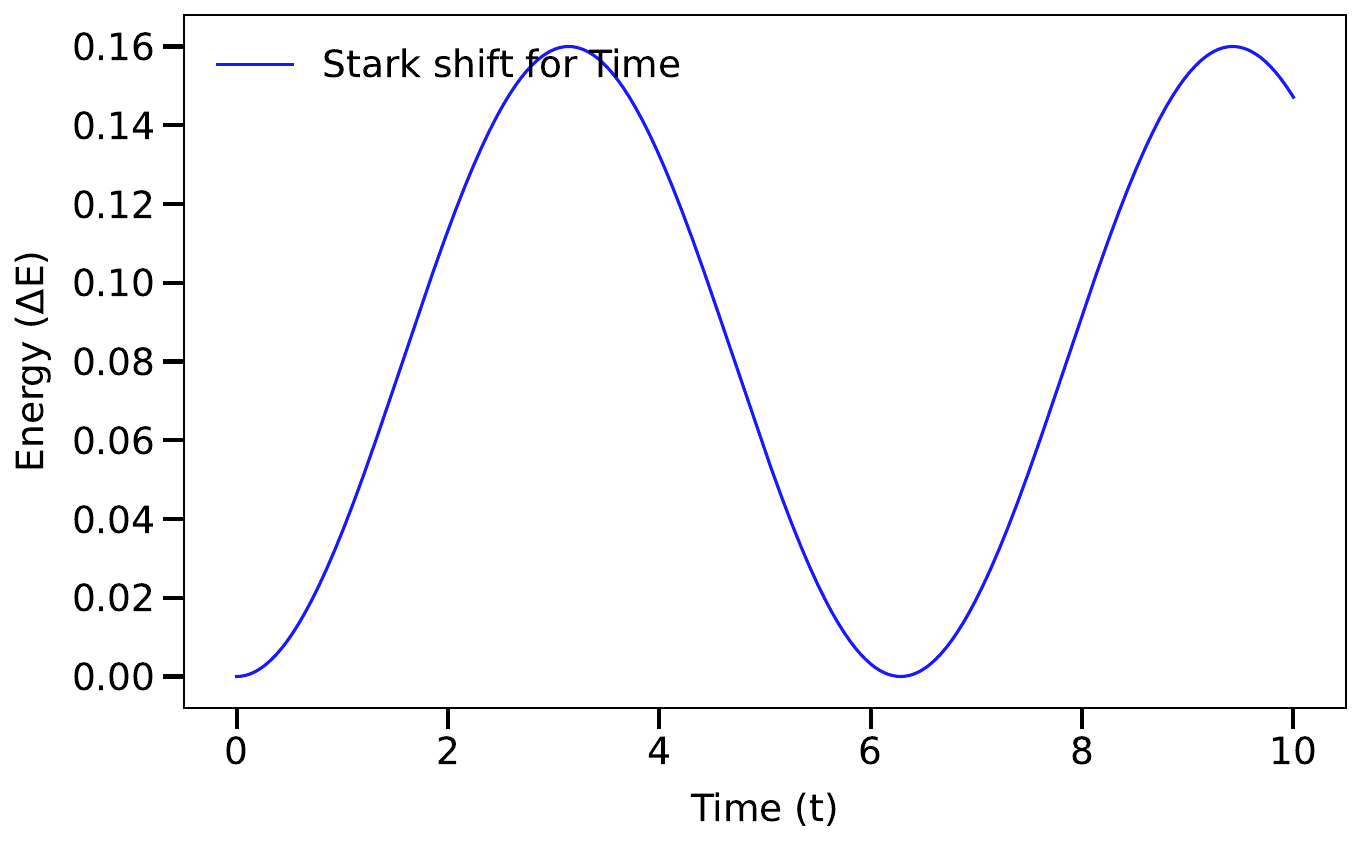}
  \caption{Variation of Stark shift with time. We choose laser parameters $\omega=0.1 \text{a.u}$, $\mathcal{E}=0.01  \text{a.u}$ }
\label{fig3}
\end{figure}

Next, we fix the frequency and vary the strength of the external laser to look into its impact on the Stark shift. This time, we predict a similar behavior, because the electron is expected to come closer to the nucleus, which experiences a greater magnitude of electric field. Thus, Stark shift should increase once again. Result has been shown in Figure \ref{fig2}.
Lastly, we keep frequency and field strength of the laser unaltered and see the splitting with passage of time. Figure \ref{fig3} illustrates the way of behaving of Stark shift which is oscillating about a fixed value. Here, laser-atom interaction time is varied from $1 fs$ to $5fs$. This is something trivial because the period Stark shift is nicely synchronized with the oscillations of the electron. Note that, by changing the laser time we mean that we have not changed the laser but we have increased the number of active cycles of the laser and the rest of the cycles are inactive or silent.

\section{Summary}\label{sec4}
Nucleus being a bound state possesses discrete energy levels as dictated by the nuclear shell model. In this paper, we study the shifts in these levels produced by the external laser, we are calling this \emph{AC nuclear Stark shift}. We impose some restrictions on the choice of laser parameters to avoid ionization and electron-nucleus collision in  H-atom. We oscillate the electron of Hydrogen using an intense laser. Due to these oscillations, the nucleus experiences a periodic electric field which is actually the source of shifting the nucleus energy levels. These levels are controlled by the varying external laser parameters. This research contributes to understanding indirect nuclear interactions in intense laser fields, laying groundwork for future experimental and theoretical advancements.

\section*{Acknowledgment}
This work is supported by the John Templeton Foundation under the grant RN049A1.

\bibliography{references}
\end{document}